\documentclass[aps,twocolumn,showpacs]{revtex4}
\usepackage{graphicx}
\begin{document}

\flushbottom

\title{Relation between space charge limited current and 
power loss in open drift tubes }
\vskip 0.3 in
\author{Debabrata Biswas}
\author{Raghwendra Kumar}

\affiliation{Theoretical Physics Division,
Bhabha Atomic Research Centre,
Mumbai 400 085, INDIA}
\email{dbiswas@barc.ernet.in, raghav@barc.ernet.in }

\date{\today}
\vskip 0.2 in
\begin{abstract}
 \noindent
Drift space is a region free from externally applied fields. It is an 
important part of many devices involving charged particle beams.
The space charge effect imposes a  limit on the current that can be transported 
through a drift space. A reasonable estimate of the space charge limited current density
($J_{SCL}^{c}$)
in {\em closed} drift tubes can be obtained from  electrostatic considerations despite the
fact that charge particle transport results in electromagnetic radiation.
We deal here with the situation where the drift tube is {\em open} to electromagnetic 
radiation, for instance due to the presence of a dielectric window.   
In such cases, electromagnetic radiation leaks out of the window which results
in a decrease in the average kinetic energy of electrons. 
If the injected current density is much lower than $J_{SCL}^{c}$,
power loss does not result in a change in the transmitted current. As the
injected current density is increased, power loss increases and at a critical 
value lower than $J_{SCL}^{c}$, reflection of electrons begins to occur.
We also show that the lowering of the space charge limited current on
account of power loss can be incorporated in the standard electrostatic
picture by introducing an effective voltage for the collection plate.

\vskip 0.25 in
\end{abstract}
\maketitle

\date{today}
\newcommand{\be}{\begin{equation}}
\newcommand{\ee}{\end{equation}}
\newcommand{\bea}{\begin{eqnarray}}
\newcommand{\eea}{\end{eqnarray}}
\newcommand{\Tbar}{{\overline{T}}}
\newcommand{\En}{{\cal E}}
\newcommand{\Lop}{{\cal L}}
\newcommand{\DB}[1]{\marginpar{\footnotesize DB: #1}}
\newcommand{\q}{\vec{q}}
\newcommand{\kt}{\tilde{k}}
\newcommand{\noi}{\noindent}

\newpage

\section{Introduction}
\label{sec:intro}

The motion of charged particles in a drift space is of considerable 
interest in fields such as high power microwaves  and accelerators 
\cite{hpm, BB}.  Here, the mutual interaction among the charged 
particles (known as the space 
charge effect) plays an important role in the dynamics of an intense beam in 
such cases \cite{BB}. 
If a beam of charged particles is injected into a drift region, space 
charge forces oppose the  incoming beam.  For a low intensity beam, space charge
effect is negligible and the transmitted current is equal to the injected  
current.  As the injected current increases to a certain critical value
(commonly referred to as the space charge limited (SCL) current), a virtual 
cathode appears in the drift space from which some of the electrons turn back to the
injecting plane \cite{BB, report}. As the injected current is
increased beyond the SCL value, the transmitted current decreases and
saturates asymptotically to a much smaller value when the 
injection energies are non-relativistic.

It is possible to estimate the space charge limited current for a 
{\em  non-relativistic} electron beam moving in an 
infinite parallel plate drift space. The SCL current density  is 
given by \cite{BB, report}

 \be
J_{\rm SCL}^{c} = {32\over9}{\epsilon_{0}}{V^{3/2} \over L^2} 
\left({2e\over m_0}\right)^{1/2}
\label{eq:SCL_1d}
\ee

\noindent      
with the injection velocity $v_{0}= \sqrt{2eV/m_{0}}$ and $L$ being the 
separation between the plates. Here  $e$ and $m_{0}$ are the electron charge  
and rest mass respectively. 
On the other hand, for a relativistic solid
beam of radius $r_{b}$ in an infinitely long drift tube of radius $R$, a 
reasonable approximation is \cite{RB}

\be
J_{\rm SCL}^{c} \simeq {m_{0}c^{3}}{4\epsilon_{0}\over e}{\left[{r_{b}^{2}}
[1+2{\rm ln}(R/r_{b})]\right]^{-1}}{\left({\gamma_{0}^{2/3}}-1 \right )^{3/2}}  
\label{eq:SCL_2d}
\ee
\noindent      
where $\gamma_{0}$ is the relativistic factor for the injected velocity.

When the injected current density $J_{IN}$ exceeds $J_{\rm SCL}^{c}$, the transmitted
current density $ J_{TR} \leq J_{\rm SCL}^{c}$ and its 
approximate analytical value can be obtained from the  so called classical 
theory in the 1-dimensional case \cite{BB,report}. 
 
It should be noted that the above estimates for the space charge limited 
current in a drift tube are derived from  electrostatic models. In reality,
the mutually interacting charged particles radiate and in the 
process, lose their kinetic energy to the fields. In closed drift
tubes, the electrostatic predictions are nevertheless good as the 
radiated electromagnetic energy is reabsorbed by the electrons.
However, when a drift tube is open to electromagnetic radiation, 
this energy leaks out as we shall demonstrate here. This leads 
to a drop in the kinetic energy of the transported electrons. 
When the injected current density, $J_{IN} <<
J_{SCL}^c$, the transmitted current $J_{TR}$ is unaffected by the
slowing down of electrons as the charge density increases to compensate
for the lower velocity. However as $J_{IN}$ increases further, reflection
begins to occur at a value (of $J_{IN}$) smaller than $J_{SCL}^{c}$.
We shall show here that the lowering of $J_{SCL}^{c}$ can be 
predicted reasonably well within the electrostatic framework 
by altering the boundary conditions to account for power loss.

It may be noted that a similar phenomenon occurs in   open diodes. 
It has recently been shown \cite{pop4} that  leakage of electromagnetic 
energy from a diode results in 
lower transmitted current. Furthermore, the electromagnetic power loss can be 
incorporated  into the standard Child-Langmuir\cite{child, langmuir} 
expression through an effective 
voltage. For a given applied potential difference and anode-cathode 
separation, the modified Child-Langmuir law can also be used to predict the 
maximum power that can leak from an open-diode \cite{pop4}.

In this paper, we present results of particle-in-cell (PIC) simulation for 
open drift tubes. In section \ref{sec:power_loss}, we
demonstrate that an opening in the drift tube allows radiation to 
leak out and lowers the kinetic energy of electrons. We next
show that the critical current density at which
reflection begins to occur in open drift tubes ($J_{SCL}^{o}$) 
is lower as compared to closed drift tubes. We also show that the
transmitted current decreases as the injected current is increased
beyond $J_{SCL}^{o}$. In section \ref{sec:eff}, we propose that 
the power loss in open drift tubes can be incorporated within 
the electrostatic framework by introducing an effective voltage for the
collection plate. The predictions of this effective energy theory
is then compared with PIC simulations. Finally, the significance 
of this study and a summary of our results are presented in the
concluding section.

\section{Power loss in Open Drift Tubes}
\label{sec:power_loss}

We shall  first demonstrate that an opening in the drift tube leads to
power loss in the form of electromagnetic radiation and hence to a
lowering of the kinetic energy of electrons reaching the far end
of the drift tube (hereafter referred to as the collection plate).

\begin{figure}[tbp]
\begin{center}
\includegraphics[width=5.5cm,angle=360]{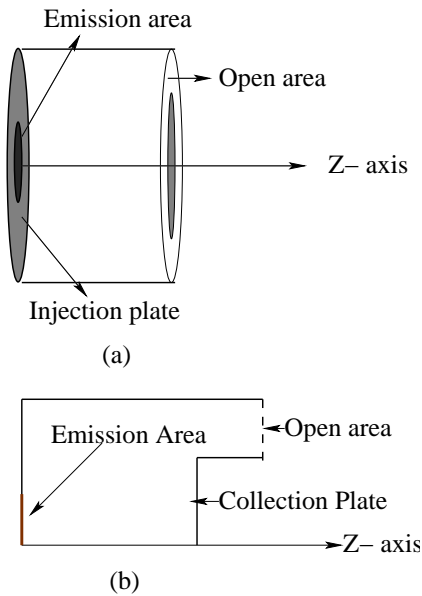}
\end{center}
\caption[ty] {Schematic of the open drift tube and the simulation geometry
as used in the PIC code XOOPIC.}
\label{fig:ODT}
\end{figure}

The drift tube under consideration is a metallic cylindrical cavity with one 
end partially open (see fig.\ref{fig:ODT}). The radius of the collection plate
($R_{CP}$) is varied in the simulation. The electrons are  injected from the 
injection plate 
in the axial direction with uniform velocity. The radius $R_{1}$ of the drift 
tube is taken to be 12.5 cm while the length  $L$ is 2.5 cm. An 
electron beam of radius $3.5$ cm is used with an injection energy of 
$50$ KeV.

The results presented here have been obtained using the fully electromagnetic
two and half dimensional PIC code XOOPIC \cite{xoopic}. The cavity wall is
considered as ``Conductor'' in the simulation; the emission model is chosen
to be ``VarWeightBeamEmitter'' while the open window is modelled by 
an ``Exitport''. 
The number of mesh points in the radial and the axial directions is typically 
64 and 128 respectively. The time step is typically taken to be 
$4 \times 10^{-13}$ s. Note that there
is no externally applied magnetic field in the simulation.

For the injection energy and plate separation under consideration,
Eq.~\ref{eq:SCL_1d} predicts the space charge limited current 
$I_{SCL}^{c} = 1283$ amperes. While this is strictly true for 
infinite parallel plates separated by a distance $L$, for the
closed  cylindrical drift tube geometry ($R_{CP} = 12.5$cm), our PIC 
simulations show that $I_{SCL} \simeq 1320$. Hereafter, we
shall refer to this value as $I_{SCL}^{c}(2)$ where the superscript
$c$ refers to the {\em  closed} drift tube and (2) refers to the
2-dimensional case.

\begin{figure}[tbp]
\begin{center}
\includegraphics[width=5.5cm,angle=270]{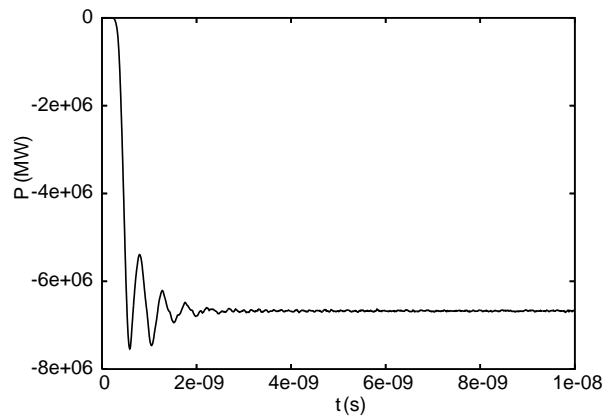}
\end{center}
\caption[ty] {The power radiated through the open area is computed using the
Poynting vector. The equilibrium value is found to be 6670 KW. }
\label{fig:power_vs_t}
\end{figure}

\begin{figure}[tbp]
\begin{center}
\includegraphics[width=5.5cm,angle=270]{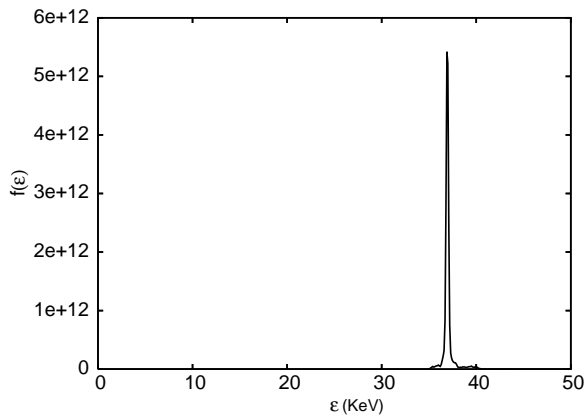}
\end{center}
\caption[ty] {The kinetic energy distribution of the electrons reaching
the collecting plate. There is a sharp peak at 36.9 KeV. }
\label{fig:energy_dis}
\end{figure}

In order to demonstrate that charge particle transport indeed 
results in electromagnetic radiation which can leak out from 
an {\em open} drift tube, we have carried out a simulation 
with the radius of the collection plate $R_{CP} = 8$ cm and the 
injected current $I_{IN} =  500$ A. The electromagnetic power
emitted from the open area is plotted in fig.~\ref{fig:power_vs_t}.
In order to account for this power loss, we also plot the
energy distribution of the electrons reaching the collection
plate in fig.~\ref{fig:energy_dis}.

As reflection of electrons does not occur at this value of 
injected current, the ratio $eP/I_{TR} \simeq e~6670/500 = 13.34$ KeV 
is the average energy loss that a single electron suffers. 
In other words, the average kinetic energy of electrons 
reaching the collection plate should be $36.66$ KeV. This
should reflect in the energy distribution of electrons at the 
collection plate. Indeed Fig.~\ref{fig:energy_dis} shows that
the kinetic energy peaks at 36.9 KeV which agrees well with the
value expected from energy conservation.

\begin{figure}[tbph]
\begin{center}
\includegraphics[width=5.5cm,angle=270]{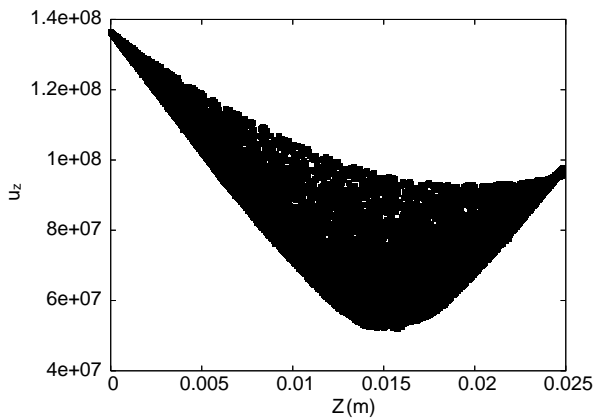}
\end{center}
\caption[ty] {The phase space for $I_{IN} = 910$ A.  There is no
reflection at this injected current.}
\label{fig:phase1}
\end{figure}

\begin{figure}[tbph]
\begin{center}
\includegraphics[width=5.5cm,angle=270]{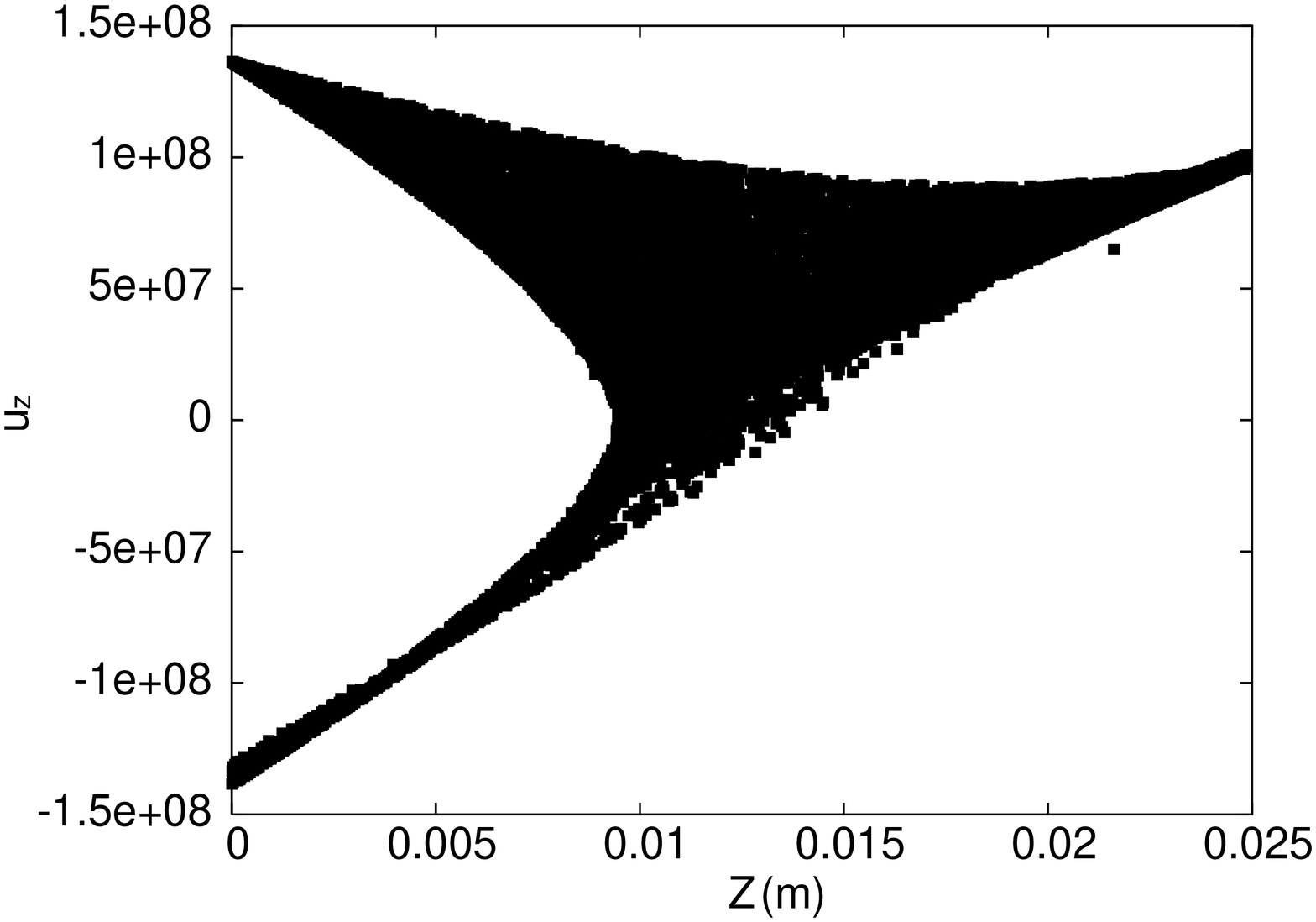}
\end{center}

\caption[ty] {The phase space for $I_{IN} = 925$ A. 
There is reflection at this injected current.}
\label{fig:phase2}
\end{figure}

\begin{figure}[tbph]
\begin{center}
\includegraphics[width=5.5cm,angle=270]{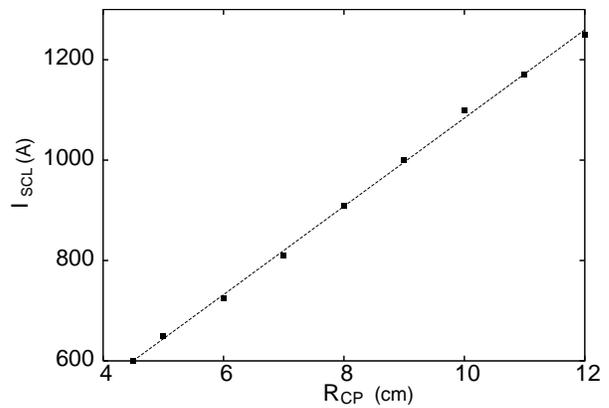}
\end{center}
\caption[ty] {The critical current as a function of the radius of 
the collection plate. The dashed line is the best fitting straight line}
\label{fig:crit_vs_opening}
\end{figure}

It is thus clear that electrons lose energy in the form of 
electromagnetic radiation and as a result slow down. This 
leads to an increase in charge density and hence enhances 
the space charge effect. For the values of injected current, energy
and power loss in the above example, the space charge 
effect is not large 
enough to cause reflection of electrons. As the injected current
is increased further, power loss increases as well thereby leading to
an increase in charge density. For the {\em open} drift tube under 
consideration, reflection occurs at $I_{SCL}^{o} \simeq 920$ as
evident from the phase space ($z,u_z=\gamma v_z$) plots at $I_{IN} = 910$ and 
$925$ amperes (see Figs.~\ref{fig:phase1} and \ref{fig:phase2}).
Thus, an opening in the drift tube lowers the space charge
limited current due to an enhancement of space charge effect caused
by the slowing down of electrons.
 
\begin{figure}[tbp]
\begin{center}
\includegraphics[width=5.5cm,angle=270]{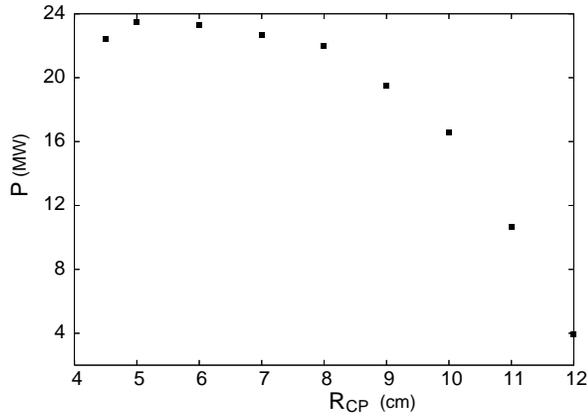}
\end{center}
\caption{Electromagnetic power emitted though the open window
is plotted against $R_{CP}$.}
\label{fig:power_vs_opening}
\end{figure}

The magnitude of power loss depends on the transmitted current
as also the size of the opening. Thus, the space charge limited
current should decrease as the size of the opening increases. 
In Fig.~\ref{fig:crit_vs_opening}, the critical current
is plotted against $R_{CP}$. For smaller openings (large $R_{CP}$),
power loss is small and thus $I_{SCL}^{o}$ is close to (but less than)
$I_{SCL}^{c}(2)$. As the size of the opening increases, power loss
increases thereby reducing the space charge limited current. 
As the opening increases further, the drop in space charge limited 
current reduces the field energy produced sufficiently so that 
power loss no longer increases as the opening is increased further. 
This is illustrated in Fig.~\ref{fig:power_vs_opening} where the 
power starts reducing at smaller openings.

For completeness, we have also studied the regime with $I_{IN} > I_{SCL}^{o}$
for three different openings. As in case of a closed drift space (1-dimensional
case) \cite{BB}, the transmitted current drops as the injected current
is increased beyond $I^{o}_{SCL}$. Note that the ``classical theory'' for 
the 1-dimensional case predicts a drop in the transmitted current 
beyond $I_{SCL}^{c}$ and a gradual decay to $I_{SCL}^{c}/8$ 
thereafter.    

\begin{figure}[tbph]
\begin{center}
\includegraphics[width=5.5cm,angle=270]{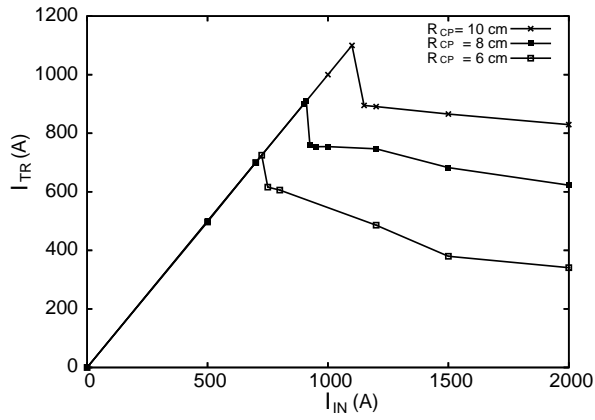}
\end{center}
\caption[ty] {The transmitted current drops as the injected current
is increased beyond $I_{SCL}^{o}$. As the opening increases, the 
space charge limited current decreases.}
\label{fig:Itr_beyondSCL}
\end{figure}

\section{The Effective Voltage Theory for Drift Tubes}
\label{sec:eff}

In the previous section we have seen that radiative power loss due
to an opening in the drift tube leads to an equivalent drop in the
electron energy reaching the collection plate. Recall that in the
standard electrostatic picture \cite{BB}, the emission and
collection plate are at the same potential $V$ where $eV$ is the 
kinetic energy of the injected electrons.
Eq.~\ref{eq:SCL_1d} follows on solving Poisson equation
and demanding that the potential minima be positive.

For open drift tubes, power loss can be incorporated within the
electrostatic picture by modifying the boundary conditions appropriately.
The loss of electron kinetic energy to electromagnetic radiation is
equivalent to having the collection plate potential at a value 
$P/I_{IN}$ lower than the emission plate potential. Thus, if the 
emission and collection plates are at $z=0$  and $z=L$ respectively,

\bea
\phi(0) & = & V  \\
\phi(L) & = & V - P/I_{IN}
\label{eq:boundary_cond1}
\eea

\noi
where $P$ is the electromagnetic power lost through the open window. 
Note that in steady state (no reflection) $I_{TR} = I_{IN}$. The 
space charge limited current for open drift space can thus be determined by 
demanding that the potential minima be positive.

Equivalently, one may study the open drift tube problem as a diode with
non-zero injection energy ${\cal E}$  \cite{pop3,akimov} having plate potentials

\bea
\phi(0) & = & 0  \\
\phi(L)& = & V_0.
\label{eq:boundary_cond2}
\eea
   
\noi
This is a well studied problem (see for instance Eq.~(19) of \cite{pop3}) for which
the space charge limited current density is

\be
J_{SCL}^{Diode} =  {J_{SCL}^{c} \over 2^3}\left( 1 + 
\sqrt{1 + {V_0 \over {\cal E} }} \right)^3
\label{eq:SCL_non0}
\ee

Applying Eq.~\ref{eq:SCL_non0} to the open
drift tube problem with $V_0 = - P/I_{IN}$, 
the space charge limited current in terms of the
radiative power loss $P$ is:

\be
I_{SCL}^{o} = {\cal A} \alpha_g {J_{SCL}^{c} \over 2^3}\left( 1 + 
\sqrt{1 - {eP \over {\cal E} I_{SCL}^{o}}} \right)^3
\label{eq:SCL_open}
\ee

\noi
where ${\cal A}$ is the emission area, ${\cal E}$ is the injection energy and 

\be
\alpha_g = J_{SCL}^{c}(2)/J_{SCL}^{c}
\ee

\noi
is a factor that accounts for the finite geometry (for geometric
factor in diodes, see \cite{lau}). In the above $J_{SCL}^{c}(2)$
is the space charge limited current in a finite closed drift tube.

Note that Eq.~\ref{eq:SCL_open} does not directly give the space charge 
limited current. Instead, for each allowed power loss $P$, one must find the
roots of Eq.~\ref{eq:SCL_open} to determine the space charge limited current $I_{SCL}^{o}$.
When $P = 0$ (closed drift tube), $J_{SCL}^{o} = \alpha_g J_{SCL}^{c} = J_{SCL}^{c}(2)$.
As the power loss increases, the number and values of allowed (positive) roots of
Eq.~\ref{eq:SCL_open} can be determined. However beyond a certain power loss, $P_{max}$,
there is no allowed root of Eq.~\ref{eq:SCL_open}. There is thus a maximum 
value of electromagnetic  power loss for a given drift tube geometry and
injection energy. 

\begin{figure}[tbp]
\begin{center}
\includegraphics[width=5.5cm,angle=270]{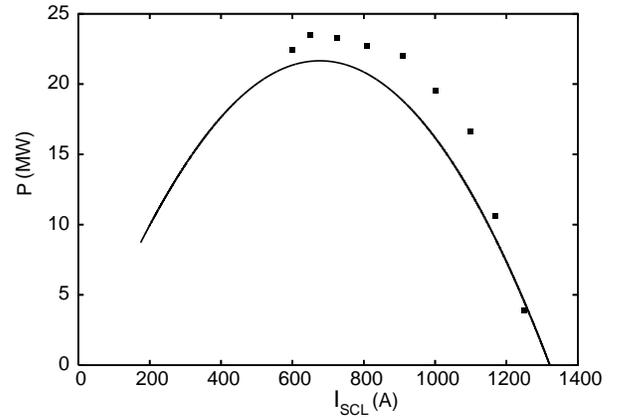}
\end{center}
\caption[ty] {The predictions of the effective voltage theory (Eq.~\ref{eq:SCL_open})
compared with PIC simulations (solid squares).}
\label{fig:eff_volt}
\end{figure}

Fig.~\ref{fig:eff_volt} illustrates these phases for the open drift
tube under consideration. The solid line represents the roots of Eq.~\ref{eq:SCL_open}
for each value of power loss $P$ and with $\alpha_g = 1.03$. 
The transmitted current is maximum when the 
drift tube is closed and $P = 0$.  Note that as $P$ increases from 0 to $P_1 \simeq 8.4$,
there is a single allowed value of $I_{SCL}$. This follows from Eq.~\ref{eq:SCL_open}
on noting that the expression within the square root must be 
positive. At $P_1$ for instance, the allowed values of current must exceed 
$I_{SCL} = 168$. For $P > P_1$, there are two admissible roots and 
these correspond to the case with (i) small opening and large current
and (ii) larger opening but smaller current. Beyond $P_{max} \simeq 21.65$,
there is no allowed current for the given geometry and injection energy.
Fig.~\ref{fig:eff_volt} also shows our PIC simulation results (solid squares)
for values of $R_{CP}$ ranging from 4.5 cm to 12 cm.
The agreement with the theoretical prediction is reasonably good.

\section{Discussion and Conclusions}

In the preceding sections, we have studied the effect of 
electromagnetic power loss on the space charge limited current in 
open drift tubes. We have demonstrated using PIC simulations that
even well below the space charge limited current, 
there is power loss which reflects in the kinetic energy of electrons reaching the 
collection plate. The slowing down of electrons however does not 
affect the transmitted current as long as the injected current is small.
As the injected current increases, the power loss increases as well
thereby leading to an enhancement of space charge effect. Thus the space
charge limit is reached earlier in open drift tubes.

We have also shown that an estimate of the space charge limited current
in open drift tubes can be obtained from electrostatic considerations
by modifying the boundary conditions appropriately using an {\em effective 
voltage} so as to reflect the loss in energy of electrons. The predictions
of this theory are in close agreement with PIC simulations.  

A fallout of this theory is the existence of a limit for the 
electromagnetic power emanating from a given open drift tube geometry and
injections energy. This can be profitably used in the design of 
devices used for the extraction of electromagnetic power \cite{eff_vir}.

 \newcommand{\PR}[1]{{Phys.\ Rep.}\/ {\bf #1}}
\newcommand{\PRL}[1]{{Phys.\ Rev.\ Lett.}\/ {\bf #1}}
\newcommand{\PP}[1]{{Phys.\ Plasmas\ }\/ {\bf #1}}



\begin{thebibliography}{99}
\bibitem{hpm}{\em High power Microwave sources}, edited by V.L. Granatstein and
I. Alexeff (Artech House,Boston,1987).


\bibitem{BB} C.K.Birdsall and W.B.Bridges, {\it Electron Dynamics of
diode regions} (Academic Press, New York, 1966).



\bibitem{report} R.~R.~Puri, R.~Kumar and D.~Biswas, {\em Theory of
Vircators 1: One dimensional model} (BARC Report: BARC/2002/I/022).

\bibitem{RB} R.~B.~Miller, {\it Introduction to the Physics of Intense Charge Particle Beams} 
(Plenum, New York, 1982).

\bibitem{pop4} D.~Biswas, R.~Kumar and  R.~R.~Puri, \PP{12}, 093102 (2005).

\bibitem{child} C.~D.~Child, Phys. Rev. Ser. 1 {\bf 32}, 492 (1911); 

\bibitem{langmuir} I.~Langmuir, Phys. Rev. {\bf 2}, 450 (1913).


\bibitem{xoopic} J.~P.~Verboncoeur, A.~B.~Langdon
and N.~T.~Gladd, Comp. Phys. Comm. {\bf 87}, 199 (1995). The code 
is available at http://ptsg.eecs.berkeley.edu/\#software;

\bibitem{pop3} R.~R.~Puri, D.~Biswas and R.~Kumar \PP{11}, 1178 (2004).

\bibitem{akimov} P.V.Akimov, H.Schamel, H.Kolinsky, A.Ya.Ender,
and V.I.Kuznetsov, Phys.Plasmas {\bf 8}, 3788 (2001).

\bibitem{lau} J.~W.~Luginsland, Y.~Y.~Lau,
R.~J.~Umstattd and J.~J.~Watrous, \PP{9}, 2371 (2002),
Y.~Y.~Lau, Phys. Rev. Lett. {\bf 87}, 278301 (2001).


\bibitem{eff_vir} D. Biswas 
and  R. Kumar, {\it Efficiency enhancement of the axial vircator} (to be published). 


\end{thebibliography}
\end{document}